\documentclass[letterpaper, onecolumn, 11pt]{IEEEtran}
\usepackage[margin=1.0in]{geometry}
\usepackage{tabulary,graphicx}
\usepackage[T1]{fontenc}
\usepackage[utf8x]{inputenc}
\usepackage{amsthm,amssymb,url,amsmath}
\usepackage[numbers,sort&compress]{natbib}
\bibpunct[, ]{[}{]}{,}{n}{,}{,}
 
\usepackage{url,multirow,morefloats,floatflt,cancel,tfrupee}
\makeatletter
\AtBeginDocument{\@ifpackageloaded{textcomp}{}{\usepackage{textcomp}}}
\makeatother
\usepackage{colortbl}
\usepackage{xcolor} 
\usepackage{pifont}
\usepackage[nointegrals]{wasysym} 
\makeatletter

\def\mcWidth#1{\csname TY@F#1\endcsname+\tabcolsep}

\def\cAlignHack{\rightskip\@flushglue\leftskip\@flushglue\parindent\z@\parfillskip\z@skip}
\def\rAlignHack{\rightskip\z@skip\leftskip\@flushglue \parindent\z@\parfillskip\z@skip}

\if@twocolumn\usepackage{dblfloatfix}\fi 
\AtBeginDocument{
\expandafter\ifx\csname eqalign\endcsname\relax
\def\eqalign#1{\null\vcenter{\def\\{\cr}\openup\jot\m@th
  \ialign{\strut$\displaystyle{##}$\hfil&$\displaystyle{{}##}$\hfil
      \crcr#1\crcr}}\,}
\fi
}

\let\lt=<
\let\gt=>
\def\processVert{\ifmmode|\else\textbar\fi}

\@ifundefined{subparagraph}{
\def\subparagraph{\@startsection{paragraph}{5}{2\parindent}{0ex plus 0.1ex minus 0.1ex}%
{0ex}{\normalfont\small\itshape}}%
}{}

\newcommand\role[1]{\unskip}
\newcommand\aucollab[1]{\unskip}
  
\@ifundefined{tsGraphicsScaleX}{\gdef\tsGraphicsScaleX{1}}{}
\@ifundefined{tsGraphicsScaleY}{\gdef\tsGraphicsScaleY{.9}}{}
\def\checkGraphicsWidth{\ifdim\Gin@nat@width>\linewidth
    \tsGraphicsScaleX\linewidth\else\Gin@nat@width\fi}

\def\checkGraphicsHeight{\ifdim\Gin@nat@height>.9\textheight
    \tsGraphicsScaleY\textheight\else\Gin@nat@height\fi}

\def\fixFloatSize#1{}%\@ifundefined{processdelayedfloats}{\setbox0=\hbox{\includegraphics{#1}}\ifnum\wd0<\columnwidth\relax\renewenvironment{figure*}{\begin{figure}}{\end{figure}}\fi}{}}
\let\ts@includegraphics\includegraphics

\def\inlinegraphic[#1]#2{{\edef\@tempa{#1}\edef\baseline@shift{\ifx\@tempa\@empty0\else#1\fi}\edef\tempZ{\the\numexpr(\numexpr(\baseline@shift*\f@size/100))}\protect\raisebox{\tempZ pt}{\ts@includegraphics{#2}}}}

\AtBeginDocument{\def\includegraphics{\@ifnextchar[{\ts@includegraphics}{\ts@includegraphics[width=\checkGraphicsWidth,height=\checkGraphicsHeight,keepaspectratio]}}}

\def\URL#1#2{\@ifundefined{href}{#2}{\href{#1}{#2}}}

\def\UrlOrds{\do\*\do\-\do\~\do\'\do\"\do\-}%
\g@addto@macro{\UrlBreaks}{\UrlOrds}
\makeatother

\usepackage{array}
\usepackage{color}
\usepackage{grffile}
\RequirePackage{afterpage}
\usepackage{epsfig}
\usepackage{here}
\usepackage{latexsym}
\usepackage{graphicx}
\usepackage{times}
\usepackage{amssymb}
\usepackage{latexsym}
\usepackage{amsmath}
\usepackage{amsthm}
\usepackage{amssymb}
\usepackage{longtable,lscape}
\usepackage{subfigure}
\usepackage{rotating}
\usepackage{multirow}

\title{Crowd Size using CommSense Instrument for COVID-19 Echo Period} 

\author{
\begin{tabular}{*{3}{>{\centering}p{.3\textwidth}}}
{\bf Santu Sardar}  & {\bf Amit K. Mishra} & {\bf Mohammed Z. A. Khan} \tabularnewline
IIT Hyderabad & University of Cape Town & IIT Hyderabad \tabularnewline
\end{tabular}
}

\usepackage{lipsum} % for filler text
\usepackage{fancyhdr}
\begin{document}
\maketitle 

\begin{abstract}
The period after the COVID-19 wave is  called the Echo-period. Estimation of crowd size in an outdoor environment is essential in the Echo-period. Making a simple and flexible working system for the same is the need of the hour. This article proposes and evaluates a non-intrusive, passive, and cost-effective solution for crowd size estimation in an outdoor environment. We call the proposed system as LTE communication infrastructure based environment sensing or LTE-CommSense. This system does not need any active signal transmission as it uses LTE transmitted signal.  So, this is a power-efficient, simple low footprint device. Importantly,  the personal identity of the people in the crowd can not be obtained using this method. First, the system uses practical data to determine whether the outdoor environment is empty or not. If not, it tries to estimate the number of people occupying the near range locality. Performance evaluation with practical data confirms the feasibility of this proposed approach. 
\end{abstract}

\section{Introduction}

With the outbreak of COVID-19 virus, we are forced to rethink visiting public areas \cite{ng,9167388,9117157,9117186}. After the current primary wave of infection improves, we will be living in a  period when we will need to run business ``almost'' as usual without a vaccine. This phase is sometimes called the Echo-period \cite{eu}. In this scenario, a device to measure the crowd size is urgently required. Crowd sensing is an active area of research because of its diverse applications.   
 
Keeping this in mind, a passive system, called CommSense (communication-based environment sensing system), is proposed here for measuring crowd size in an outdoor environment. The concept of CommSense was proposed and  verified by the authors in simulation and with field-collected data for various indoor applications like indoor object detection, indoor localization and indoor occupancy estimation \cite{1734246}. The system was also tested for outdoor vehicle detection and classification \cite{8700094}. These give the confidence to apply the CommSense principle to perform outdoor crowd size measurement. It can be noted that a major advantage of the CommSense system, compared to audio or vision-based methods, is that our system is non-intrusive. It cannot identify the persons. 

To demonstrate the objective, the experiments and analyses performed here are as follows: 
\begin{enumerate}
\item First, we considered that the people in the outdoor crowd are static. Crowd detection using a threshold-based method was performed first. If a crowd is detected, it triggers the investigation of the deriving number of persons in the crowd.
\item In the next step, the same analysis was repeated when the people in the crowd were moving freely.
\item Finally, we have performed these two experiments on a different day and time and compared the performance to examine the consistency of this approach.
\end{enumerate} 

\section{The State of the Art}

As crowd sensing is vital for many other applications \cite{7539244,7684720}, there have been attempts to perform outdoor crowd size estimation in the past. Video based crowd size estimation was proposed in \cite{5685978}. The performance of video processing for crowd-monitoring applications was analyzed in \cite{8355170}. A crowd counting method was proposed in \cite{6115886}, which uses detection flow along the temporal video sequence. Automated video analytic was used for crowd monitoring and counting in \cite{8926351}. Counting the number of people present in a crowd with a real-time network of image sensors was proposed in \cite{1238325}. A motion-based crowd density estimation method was proposed in \cite{8732710}. A crowd counting method based on image processing and Convolutional Neural Network (CNN) was proposed in \cite{8755826}. The proposal was verified using $UCF-CC-50$ dataset and the $ShanghaiTech$ crowd monitoring dataset \cite{8755826}. But these vision-based approaches require the crowd to be in the line-of-sight of the cameras. Their performance also depends on the overlapping of the objects, the relative positioning of the crowd, visibility at the time of the day, weather conditions, pollution level, etc. Additionally, it also poses privacy concerns. 
The authors of \cite{8117189} tried to estimate the size of dense crowds i.e., having hundreds to thousands of people present in the crowd using a distributed protocol that relies on mobile device-to-device communication. This required active transmission at all times and was verified in simulation. In addition, for the post-COVID scenario, we need crowd size estimation on a smaller scale. The work in \cite{7264373} uses smartphones’ acoustic sensors in the presence of human conversation, and motion sensors in the absence of any conversational data, for crowd size estimation. The presence of up to ten occupants was tested with this proposed system. There has been few attempts to use LTE based systems for crowd density estimation \cite{8876595,7997194}. In \cite{9085930}, the authors presented an Internet of Medical Things enabled wearable called EasyBand for autocontact tracing to limit the growth of new positive cases. Currently, the world is almost coming to a halt to reduce COVID-19 spread. Official recommendations for social distancing have pushed people into ever-smaller clusters \cite{ng}. Though the risk is low for a small event or crowd, each additional case or event of any size increases the overall risk.

\section{LTE-CommSense System}

\begin{figure}[!h]
\centering
\subfigure[]{\psfig{file=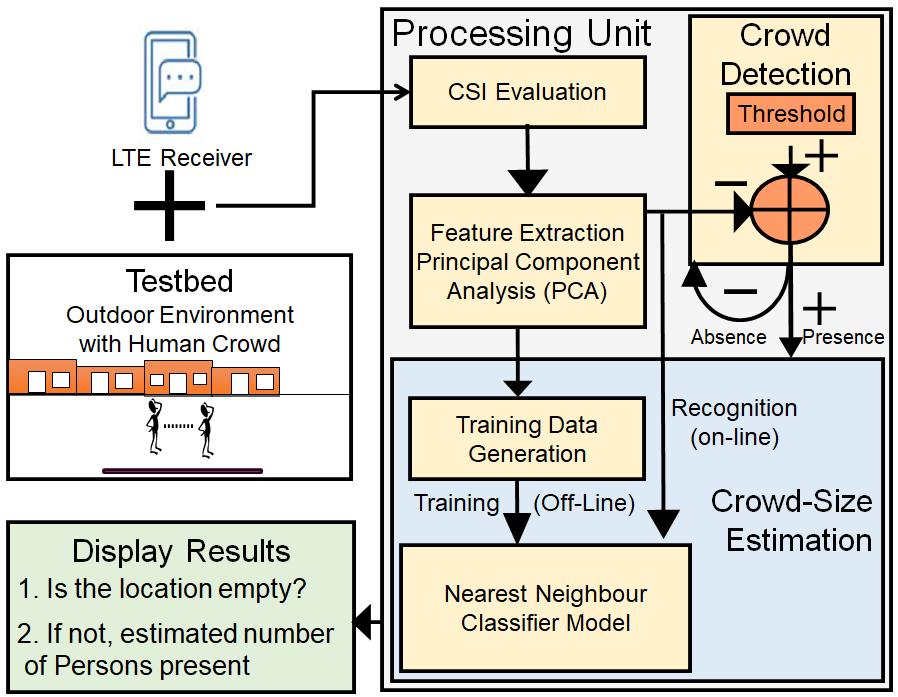,width=0.49\textwidth}}
\subfigure[]{\psfig{file=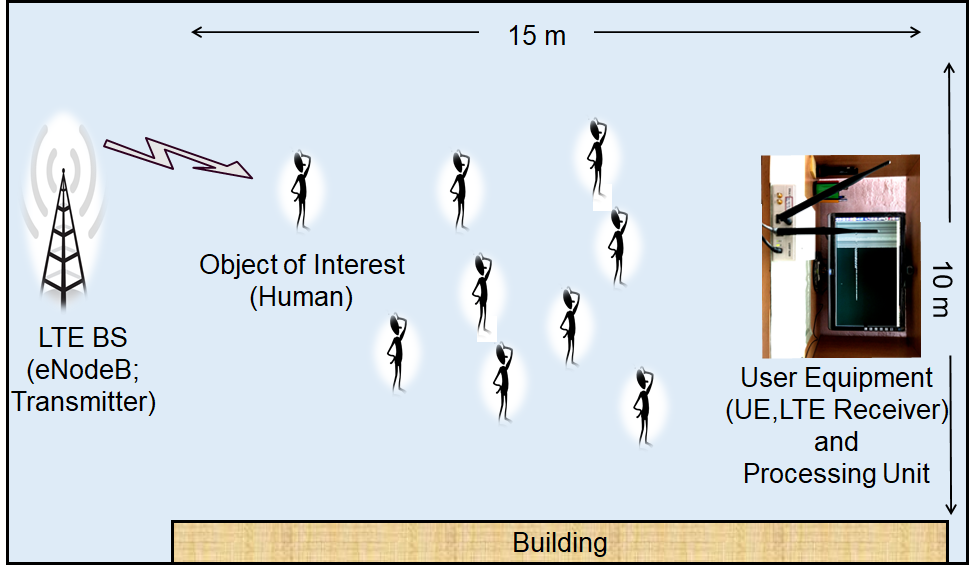,width=0.49\textwidth}} 
\caption{(a) A Block Diagram of LTE-CommSense System which uses a Single UE for Outdoor Crowd-size Estimation, (b) Relative Position of LTE receiver, Crowd and LTE eNodeB in the Outdoor Environment.}
\label{usrpn200}
\end{figure}

Fig. \ref{usrpn200}(a) shows a simple version of the proposed system which uses a piece of single user equipment (UE) \cite{1734246,8700094}. In this work, a single UE was considered for verification using practical data. The LTE UE receiver works at Band-40 (2300 MHz to 2400 MHz frequency band) in time division duplexing (TDD) topology. The signal captured had a bandwidth of 10 MHz. 

In the experimental setup, we have used a total of 19 number of people in an outdoor roadside environment. The CommSense prototype was employed to collect LTE downlink (DL) data affected by the presence of the crowd. The modelled UE, which is a part of the proposed prototype, then receives the DL data and performs standard UE operations to evaluate state information (CSI). Further processing was performed on the CSI for the detection of the crowd and onward estimation of the number of people present in the crowd. First, feature extraction was performed, and then crowd presence was detected using a threshold-based method. In case a crowd is detected, the next stage of operation initiates wherein the prototype tries to determine the number of persons present in the crowd.  

The description and working principle of an USRP N200 SDR platform can be obtained from \cite{1734246,8700094}. The SDR was modelled as the CommSense prototype containing LTE receiver. The LTE downlink signal was captured via the antenna and radio frequency (RF) daughter card connected to the N200 platform. This captured downlink data was transferred to the processing unit using Ethernet serial port interface for further computation and analysis. Fig. \ref{usrpn200}(b) explains the relative position of the prototype, crowd, and LTE base station (eNodeB) in the outdoor environment in the experimental setup. Firstly, the LTE downlink data were recorded without the presence of the crowd, and then data were captured for different numbers of people present. The size of the crowd i.e. the number of persons present in the crowd is varied from one to nineteen at an interval of three. This method is performed for two scenarios. In one case, the persons in the crowd are static, and in another case, they were moving without any restriction within the premises. For each downlink data capture, a thousand CSI values were extracted from the LTE receiver inside our prototype.

\section{A Case Study Design}

\subsection{Detection for Static and Dynamic Crowd}

As per the experimental setup, there are a total of eight different cases. The first class corresponds to the outdoor environment with the absence of any crowd. The next category denotes the presence of only one person. Category three is for the presence of four persons. In a similar manner, we keep on increasing the number of persons by three until the eighth category corresponding to the presence of nineteen persons was reached. For all these eight categories, 1000 CSI values were extracted. Now, to detect the presence of a person, Principal component analysis (PCA) was performed for feature extraction of the CSI values \cite{1734246}. 

\begin{figure}[!h]
\centering
\subfigure[]{\psfig{file=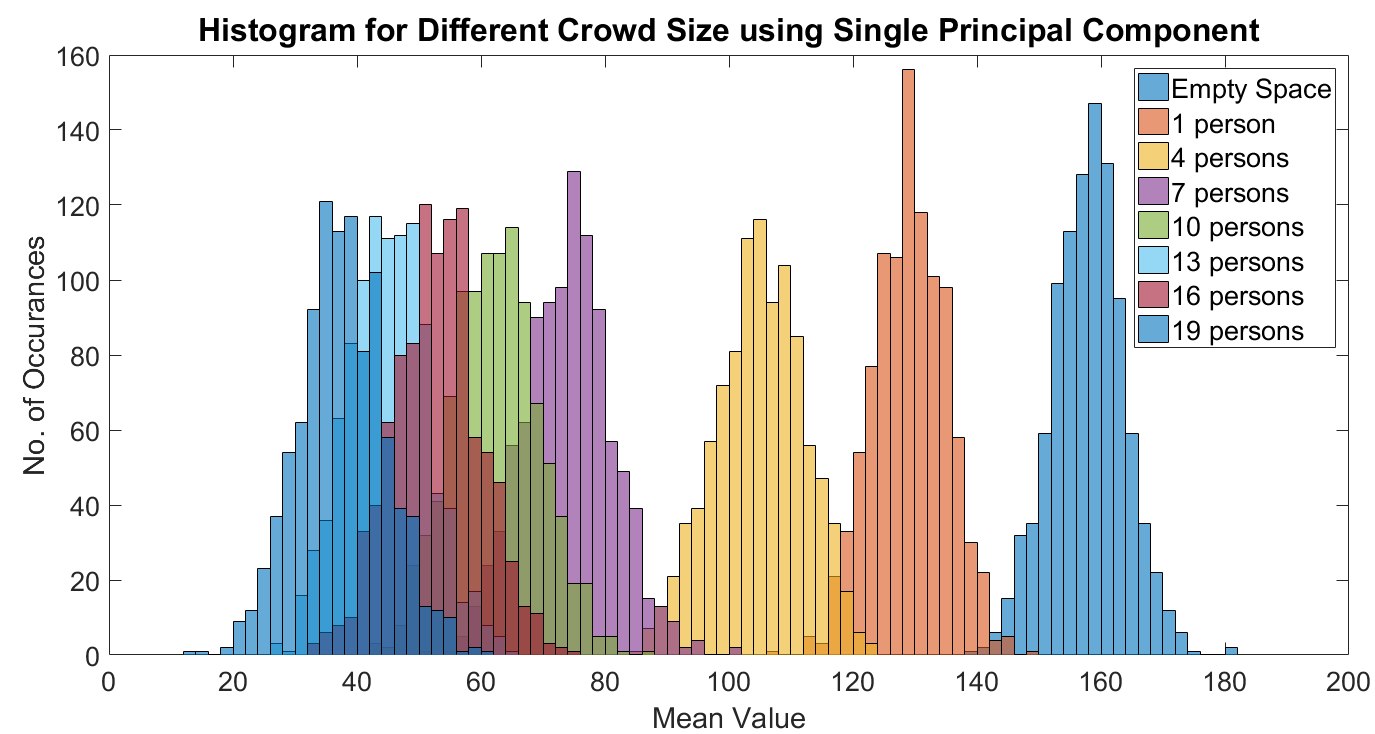,width=0.49\textwidth}}
\subfigure[]{\psfig{file=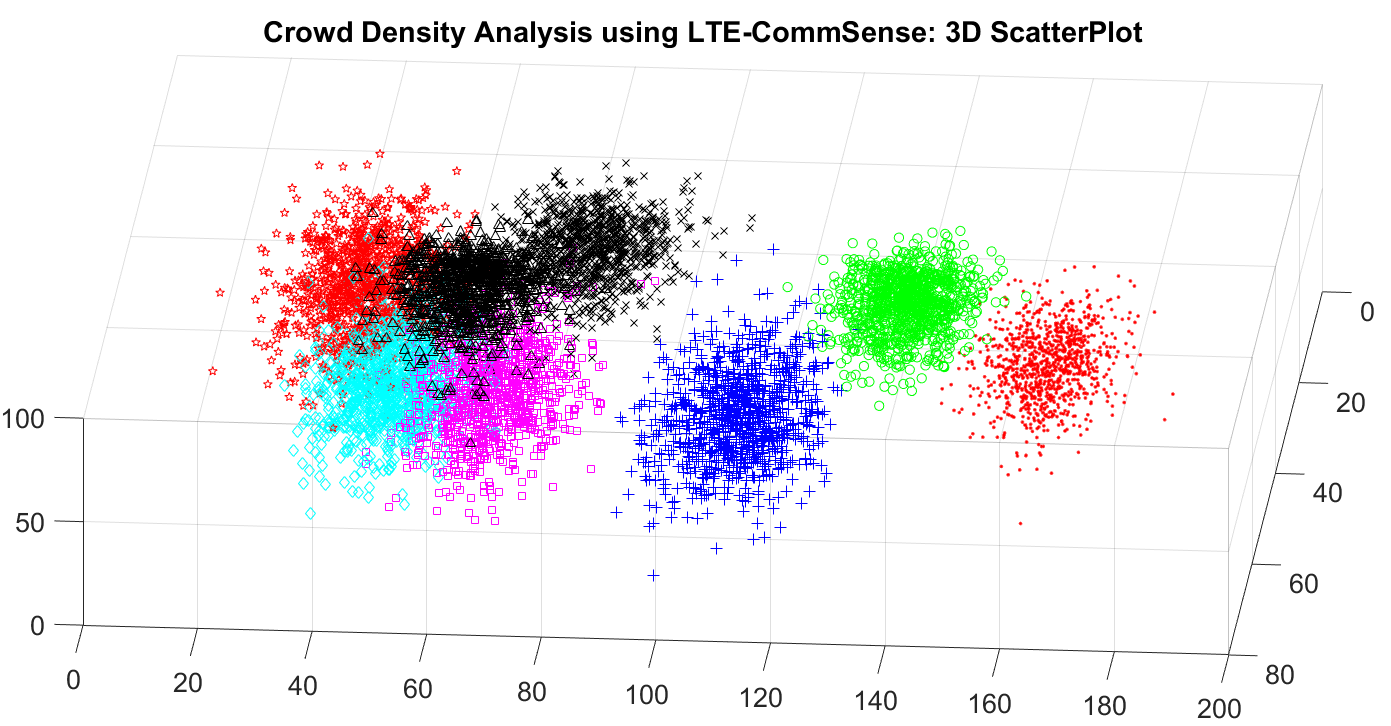,width=0.49\textwidth}}
\caption{Static Crowd Detection: (a) Histogram of Principal Component for Largest Eigenvalue, (b) Scatterplot corresponding to Highest Three Eigenvalues}
\label{static}
\end{figure}

Principal components for all the categories corresponding to the largest eigenvalue are plotted in Fig. \ref{static}(a). Fig. \ref{static}(b) depict the three-dimensional scatterplots of the principal components for the largest three eigenvalues. The description of the categories for Fig. \ref{static}(b) are explained in Table \ref{cluster_color}. In this plot, distinguishable clusters are visible corresponding to the absence of crowd and the presence of different sizes of the crowd. In Fig. \ref{static}(a), for the case where only one principal component corresponding to the largest eigenvalue was considered, the overlap between the clusters was more, and this overlap decreases gradually when more number of principal components were considered. 

\begin{table}[!h]
\caption{Category Description in the Three Dimensional Clusterplot for Three Principal Components corresponding to Highest Three Eigenvalues}
\begin{center}
\begin{tabular}{|m{10mm}|m{45mm}| m{12mm}|}\hline
Symbol & Experiment Description  & Category  \tabularnewline    \hline
{\Large \textcolor{red} {\bf \textbullet} }             & Empty Outdoor Environment  &     1    \tabularnewline    \hline
${\Large \textcolor{green} {\bf \bigcirc } }$           & No. of Persons = 1   &     2  \tabularnewline    \hline
{\Large \textcolor{blue} {\bf +}}                       & No. of Persons = 4   &     3  \tabularnewline    \hline
{\large \textcolor{black} {\bf X}}                      & No. of Persons = 7   &     4  \tabularnewline    \hline
${\Large \textcolor{magenta} {\bf \Huge \square}} $     & No. of Persons = 10   &     5  \tabularnewline    \hline
${\Large \textcolor{cyan} {\bf \Huge \diamondsuit}} $   & No. of Persons = 13   &     6  \tabularnewline    \hline
${\Large \textcolor{black} {\bf \Delta}}$               & No. of Persons = 16   &     7  \tabularnewline    \hline
${\Large \textcolor{red} {\bf \bigstar}} $              & No. of Persons = 19   &     8  \tabularnewline    \hline
\end{tabular}
\end{center}
\label{cluster_color}
\end{table}

\begin{figure}[!h]
\centering
\subfigure[]{\psfig{file=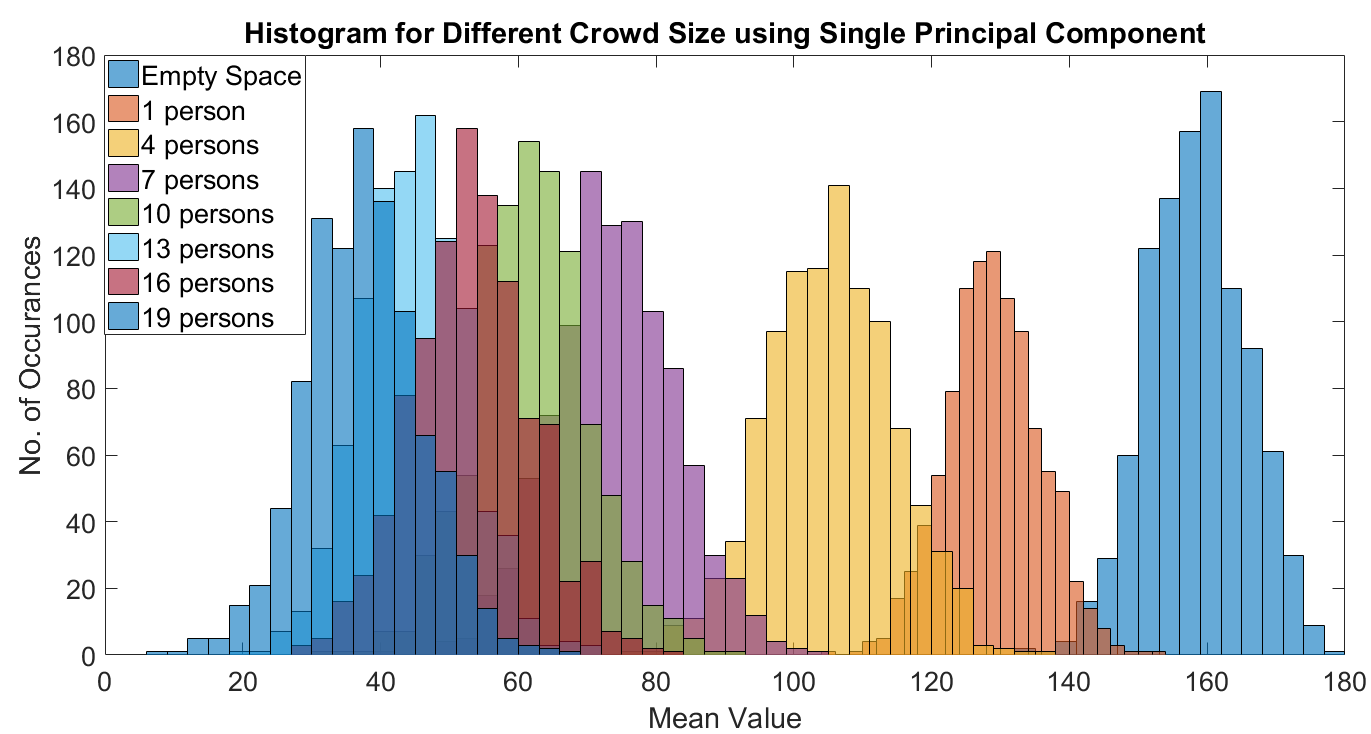,width=0.49\textwidth}}
\subfigure[]{\psfig{file=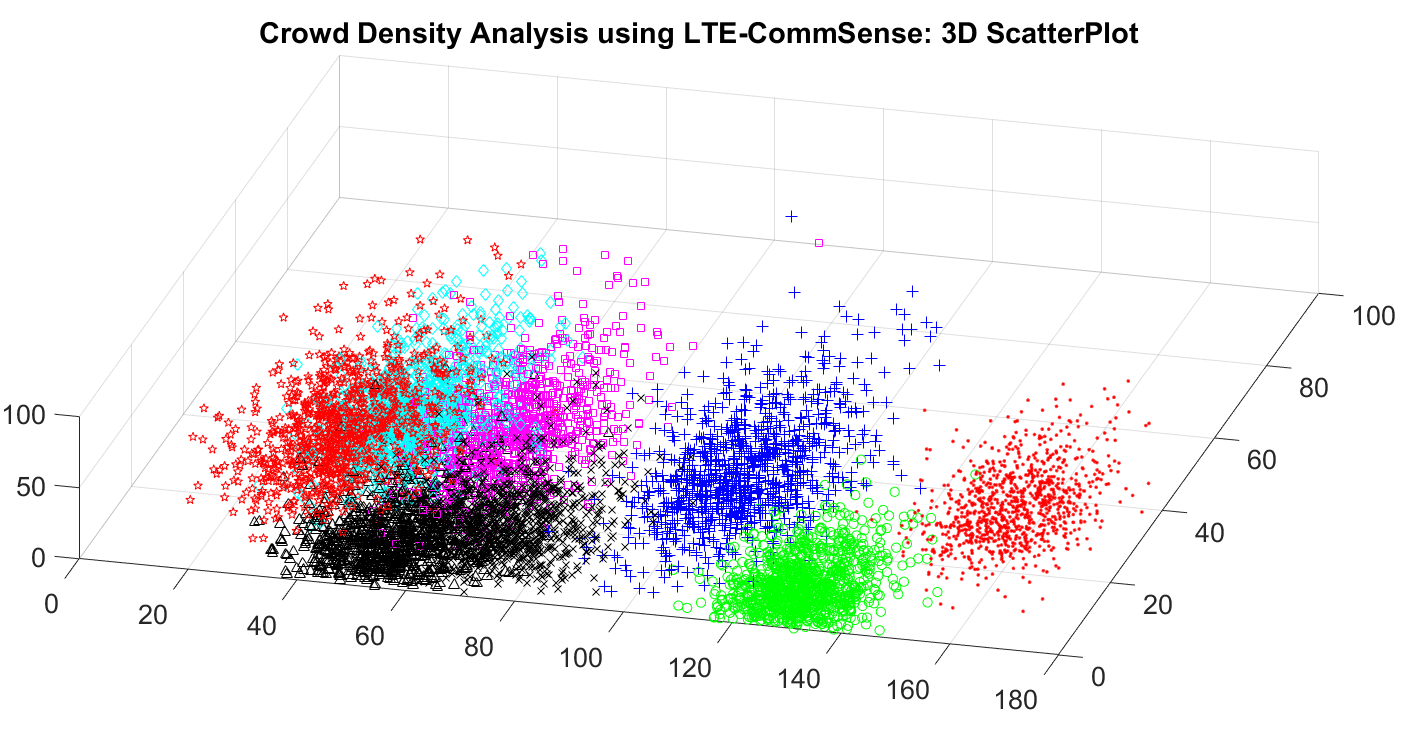,width=0.49\textwidth}}
\caption{Dynamic Crowd Detection: (a) Histogram of Principal Component for Largest Eigenvalue, (b) Scatterplot corresponding to Highest Three Eigenvalues}
\label{dynamic}
\end{figure}

The above analysis was performed next when the persons were in random motion in the locality. Principal components for all the categories corresponding to the largest eigenvalue are plotted in Fig. \ref{dynamic}(a). Fig. \ref{dynamic}(b) depict the three-dimensional scatterplots of the principal components for the largest three eigenvalues. The description of the categories for Fig. \ref{dynamic}(b) are same as in Table \ref{cluster_color}. In this plot, distinguishable clusters are visible corresponding to the absence of crowd and the presence of different sizes of the crowd. In Fig. \ref{dynamic}(a) i.e., for the case where only one principal component corresponding to the largest eigenvalue was considered, the overlap between the clusters was more, and this overlap decreases gradually when more number of principal components were considered. 

\subsection{Threshold-based Detection and Verification of Consistency}

\begin{figure}[!h] 
\centering
\psfig{file=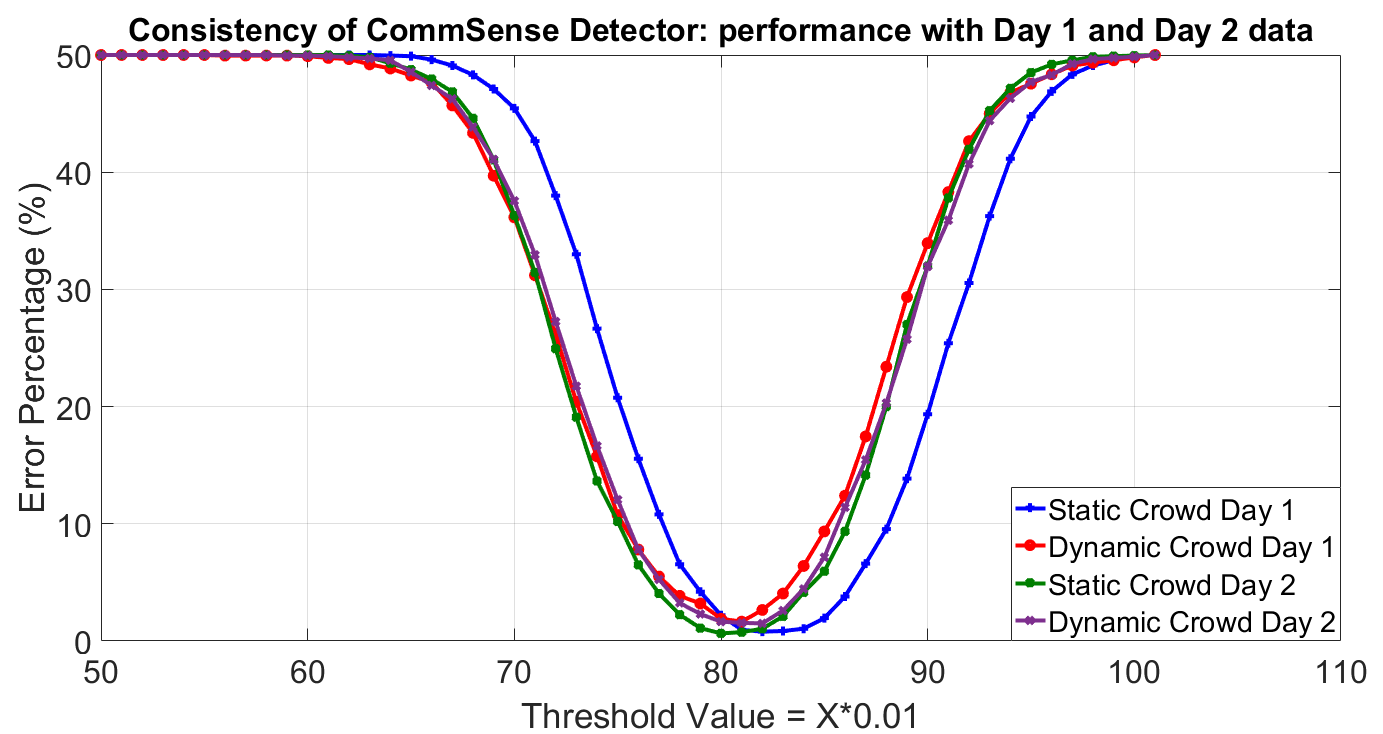,width=0.6\textwidth}
\caption{Consistency of CommSense Detector performance: Comparison of Detection Performance at Two Different Days.}
\label{allcsi}
\end{figure}

Threshold-based detection performance corresponding to different values of the selected threshold may be analyzed for both static and dynamic case. For static crowd as shown in Fig. \ref{static}(a), the error percentage of crowd detection vs. the selected threshold was shown in Fig. \ref{allcsi}. The percentage of error in detection was obtained to be $0.65\%$ for a threshold value of $0.80$. The same analysis was performed for the dynamic crowd shown in Fig. \ref{dynamic}(a). The error percentage of $1.7\%$ was evaluated for a threshold value of $0.82$ (Fig. \ref{static}(d)). Both these results are depicted in Fig. \ref{allcsi} labelled as `Static Crowd Day 1' and `Dynamic Crowd Day 1' respectively. 

We evaluated the performance consistency of the proposed approach to ascertain the reliability of the developed prototype. The exercise performed above i.e., detection of the crowd and subsequent crowd size estimation for the static and dynamic crowd, was repeated on a different day with a new set of people. In Fig. \ref{allcsi}, we can see the detection performance for the other day (Day 2) was also along with the initial day data. In `Day 2', for these two different types of crowds, minimum detection errors achieved a minimum value of $0.8\%$ and $1.55\%$, respectively. The threshold value for which the error percentages are minimum can be finalized as the threshold value with which the future data samples may be compared for detection of crowd. Comparison with the performances of Day 1 concludes that both the days' detection performance was consistent.

\subsection{Estimation of Crowd Size for Static and Dynamic Crowd}

Detection of the crowd triggers the next stage of analysis. Here, we attempt to estimate crowd size i.e. the number of persons in the crowd. This estimation was also performed for both static and dynamic conditions of the crowd. Table \ref{accu_static_dynamic} depicts the evaluated accuracy for different numbers data for training and size estimation. A simple nearest neighbor classifier was chosen for this purpose. When a new sample data is input to the system, using the nearest neighbor classifier the distance of that from the nearest data available in the training data-set was evaluated. The results reveal that proposed approach is a promising candidate for crowd detection and its size estimation in an outdoor environment consisting of either static or dynamic persons present in the crowd. Table \ref{confu} shows the confusion matrix for the lowest accuracy achieved in the Table \ref{accu_static_dynamic}. The corresponding accuracy achieved for the static and dynamic crowd cases are $86.02\%$ and $84.55\%$, respectively.

\begin{table}[!h]
\caption{Classification of Number of Persons Present for Static and Dynamic Crowd Scenario}
\begin{center}
\begin{tabular}{|m{10mm}|m{10mm}|m{15mm}|m{15mm}|}   \hline
Training Images   & Testing Images  &  Accuracy (\%) for Static Crowd & Accuracy (\%) for Dynamic Crowd \\ \hline
200      &    250  & 85.9      & 84.35       \\ \hline                                      
300      &    350  & 86.3      & 85.36       \\ \hline                                      
400      &    500  & 86.02     & 84.55       \\ \hline                                      
500      &    700  & 85.3      & 83.02       \\ \hline                                      
600      &    800  & 85.6      & 83.72       \\ \hline                                      
750      &   1000  & 85.6      & 83.91       \\ \hline                        
\end{tabular}
\end{center}
\label{accu_static_dynamic}
\end{table}

\begin{table}[!h]
\caption{Confusion Matrix for Static and Dynamic Crowd for the case of 500 Training Data per Category and 700 Testing Data per Category} %\vspace{-9pt}
\begin{center}
{
\begin{tabular}{|m{5mm} m{5mm} m{5mm} m{5mm} m{5mm} m{5mm} m{5mm} m{5mm} |}   \hline
\multicolumn{8}{|c|}{Static Crowd} \\ \hline
\multirow{8}{*}{} 697  &     3  &     0   &    0   &    0   &    0   &    0  &    0 \\
					1  &   698  &     1   &    0   &    0   &    0   &    0  &    0 \\
					0  &     1  &   699   &    0   &    0   &    0   &    0  &    0 \\   
					0  &     0  &     0   &  682   &    7   &    0   &   11  &    0 \\
					0  &     0  &     0   &    7   &  671   &   15   &    5  &    2 \\ 
					0  &     0  &     0   &    0   &   29   &  644   &    1  &   26 \\
					0  &     0  &     0   &    8   &    7   &    0   &  685  &    0 \\
					0  &     0  &     0   &    1   &    2   &   26   &    1  &  670 \\
\hline
\multicolumn{8}{|c|}{Dynamic Crowd}\\ \hline
 \multirow{8}{*}{}697  &     3   &    0  &     0   &    0   &    0   &    0   &     0\\
					1  &   685   &   14  &     0   &    0   &    0   &    0   &     0\\
					0  &    10   &  688  &     2   &    0   &    0   &    0   &     0\\   
					0  &     0   &    3  &   659   &   22   &    1   &   14   &     1\\
					0  &     0   &    5  &    12   &  639   &   28   &   14   &     2\\ 
					0  &     0   &    0  &     0   &   24   &  623   &    5   &    48\\
					0  &     0   &    0  &    11   &   11   &   10   &  658   &    10\\
					0  &     0   &    0  &     0   &    1   &   44   &    5   &   650\\
\hline
\end{tabular}
}
\end{center}
\label{confu}
\end{table}

\section{Conclusion}

Crowd detection and its size estimation in an outdoor environment are very helpful for safety in the post-COVID world. This article proposes a passive non-intrusive solution for outdoor crowd detection and subsequently, its size estimation. %If presence of crowd is detected in the outdoor location, it estimates the number of people present in the crowd. 
The feasibility of this novel approach was verified with practical signal captured using SDR based prototype developed by the authors. The detection and subsequent estimation analysis was performed for crowd in static condition. Later the same analysis was performed for the case where the persons were in motion without any restriction in the outdoor location. The results prove the feasibility of our proposal. For the static crowd case, the percentage of error in detection was $0.65\%$, whereas, for the dynamic scenario, the same was evaluated to be $1.7\%$.

Consistency of the performance of this proposal was evaluated by calculating the detection accuracy for static and dynamic crowd on a different day. The crowd was constituted with a different set of people. When compared with the initial outcomes, similar performance was observed which proves the performance consistency of the proposed method.

For the static as well as dynamic crowd cases, the nearest neighbor classifier provided acceptable performance. Fig. \ref{static}(a) and Fig. \ref{dynamic}(a) reveals that confusion is more when we try to distinguish between more number of persons. 

The analysis in this article with practical data confirms that LTE-CommSense principle can successfully detect crowd in outdoor environment. After detection, it can estimate the crowd size as well with reasonable accuracy.  

\bibliographystyle{IEEEtran}
\bibliography{paper}

\vspace{-60pt}
\begin{IEEEbiographynophoto}{Santu Sardar} received the M.Tech. degree from IIT Guwahati, India in 2010. He is pursuing his Ph.D. from IIT Hyderabad, India and heading a DRDO laboratory working on Quantum technologies. Contact him at sardar.santu@gmail.com.
\end{IEEEbiographynophoto}
\vspace{-60pt}
\begin{IEEEbiographynophoto}{Amit K. Mishra} joined IIT Guwahati, as an Assistant Professor in 2006. He is currently with the University of Cape Town, Cape Town, South Africa. Mr. Mishra is a Y-Rated Researcher in South Africa. He was a recipient of the IRSI Young Scientist Award in 2008 and the Endeavor Research Fellowship in 2010. Contact him at akmishra@ieee.org. 
\end{IEEEbiographynophoto} 
\vspace{-60pt} 
\begin{IEEEbiographynophoto}{Mohammed Z. A. Khan} has worked with Sasken, Silica Semiconductors, and Hellosoft. He is currently a Professor at  IIT Hyderabad, India. His contributions in space–time block codes are adopted by the WiMAX Standard. He was a recipient of the INAE Young Engineer Award in 2006 and the Visvesvarya Young Faculty Research Fellowship since 2016. Contact him at zafar@iith.ac.in.
\end{IEEEbiographynophoto}

\end{document}